\documentclass[letterpaper]{article} 
\usepackage{aaai2027}  
\usepackage[hyphens]{url}  
\usepackage{graphicx} 
\usepackage{natbib}  
\usepackage{caption} 
\usepackage{algorithm}
\usepackage{algorithmic}

\usepackage{newfloat}
\usepackage{listings}
\DeclareCaptionStyle{ruled}{labelfont=normalfont,labelsep=colon,strut=off} 
\floatstyle{ruled}
\newfloat{listing}{tb}{lst}{}
\floatname{listing}{Listing}

\usepackage{booktabs}

\title{Executable Specifications as First-Class Independent Verification and Validation Artifacts for Quantum Information and Machine Learning}
\author{
    Ragib Shahariar Ayon,
    Shibbir Ahmed
}

\affiliations{
    Department of Computer Science, Texas State University\\
    San Marcos, TX, USA\\
    ragibsayon@txstate.edu, shibbir@txstate.edu
}

\begin{document}

\maketitle

\begin{abstract}
Executable specifications make software assumptions explicit and checkable, yet their role in quantum information and machine learning (QIML) remains underexplored. We argue that executable specifications should be treated as first-class artifacts for independent verification and validation (IV\&V) of QIML systems. Static analysis, testing, simulation, and runtime checking can provide evidence about QIML behavior, while Design by Contract shows that selected quantum-state, measurement, and statistical properties can be expressed as checkable conditions. However, adapting existing methods for specification inference and assessment from conventional and machine-learning software to QIML raises several challenges, notably limited evidence for deriving generalizable specifications, cross-stage and probabilistic behavior, repository-scale reasoning across hybrid workflows, and weak behavioral oracles. We characterize these challenges and identify research directions for QIML corpora, specification semantics, repository-level analysis, and specification assessment.
\end{abstract}
\section{Introduction}
\label{sec:introduction}
Quantum Information and Machine Learning (QIML) applications combine classical data processing, parameterized quantum circuits, measurement, optimization, and model evaluation. Their correctness, therefore, depends on more than whether individual functions execute successfully. Assumptions about input dimensions, parameter domains, circuit construction, quantum states, measurement behavior, execution order, and classical-quantum interactions may all determine whether an execution is meaningful. When these assumptions remain implicit, a QIML program can execute without exposing the violation of an intended behavioral condition. Quantum software already provides several mechanisms for detecting problematic behavior. LintQ identifies quantum-specific programming problems through static analysis~\cite{lintq}; MorphQ uses metamorphic testing when conventional output oracles are unavailable~\cite{morphq}; and QuraTest incorporates quantum-specific properties into test generation~\cite{quratest}. Quantum Design by Contract (DbC) further shows that properties of circuit states, measurements, and statistical processing can be expressed as explicit assertions~\cite{yamaguchi2023design,yamaguchi2025practical}. Beyond runtime-oriented assertions, quantum Hoare logic provides a deductive foundation for specifying quantum preconditions and postconditions, and for reasoning about the partial and total correctness of quantum programs~\cite{ying2011floyd}.

\begin{figure}[t]
    \centering
    \includegraphics[width=\linewidth]{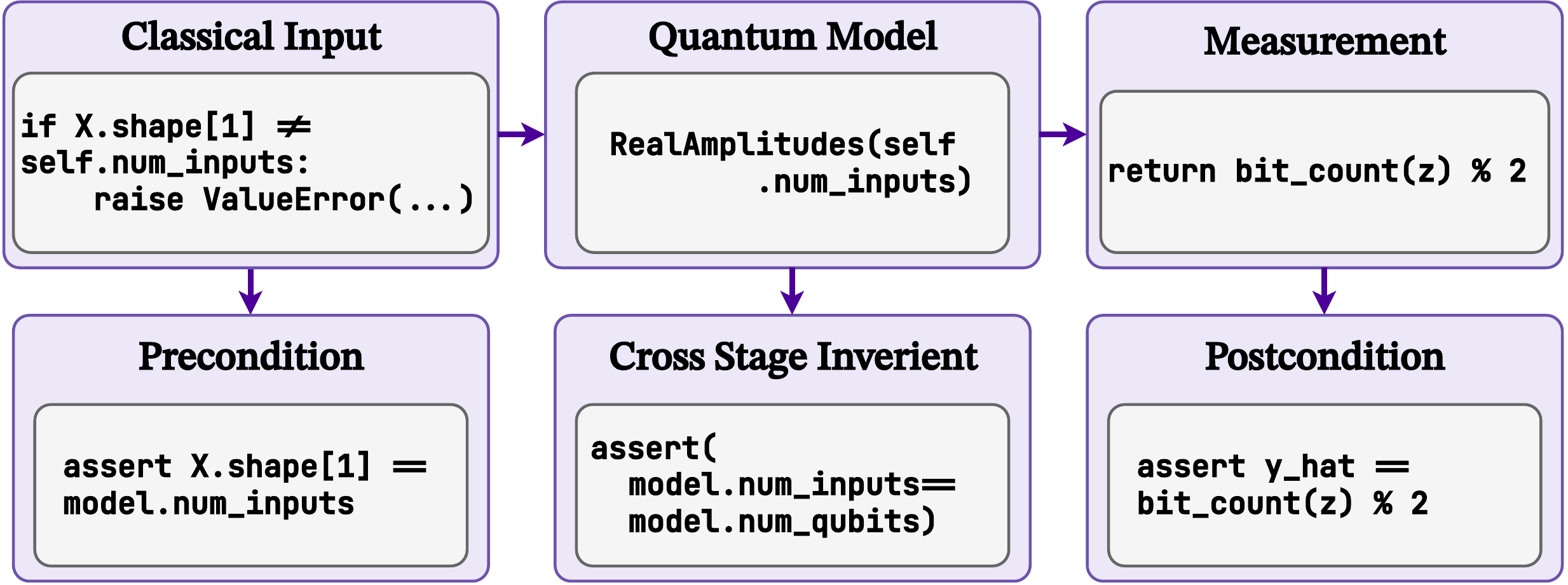}
    \caption{Illustrative Qiskit Machine Learning classifier and its executable specifications. The example exposes a classical-input precondition, a cross-stage invariant relating classical and quantum dimensions, and a postcondition on the interpretation of measurements.}
    \label{fig:qiml-worked-example}
\end{figure}

Figure~\ref{fig:qiml-worked-example} makes the underlying specification problem concrete. The classifier checks that the classical feature dimension matches the model input dimension, uses that input dimension to configure the parameterized quantum circuit, and maps measurement outcomes to a binary output. These relationships can be expressed as a precondition on the classical input, a cross-stage invariant relating the classical and quantum dimensions, and a postcondition on measurement interpretation. The example illustrates that relevant behavioral assumptions may be distributed across multiple stages of a QIML workflow rather than contained within a single function. Existing analysis and testing mechanisms address how such behavior can be checked, but a different question remains: \textit{what behavioral claim should those mechanisms evaluate?} In deep-learning software, useful specifications already cover properties of data, model structure, and execution behavior~\cite{shibbir2023dlcontract,dlpreconditions}. QIML adds cross-stage dependencies: data preparation may constrain circuit construction, parameterization may affect measurement behavior, and measurement results may guide downstream classical optimization. Thus, a behavioral obligation may be established in one stage but become observable only in another, making function-local checks insufficient to capture it. IV\&V therefore requires two separate things: a mechanism for checking the system and an explicit claim about the behavior that the mechanism is intended to check. Recent efforts such as DARPA's Quantum Benchmarking Initiative underscore the growing importance of independent assessment for emerging quantum systems~\cite{darpa_qbi_ivv_2026}. Static analysis, testing, simulation, and runtime checking can provide evidence about a system, but that evidence is informative only when the expected behavior is stated explicitly. A passing test, for example, shows that particular executions succeeded; it does not establish that all relevant assumptions or cross-stage obligations were satisfied. At the QIML software layer, IV\&V therefore requires behavioral expectations that can be stated and inspected independently of the mechanism used to check them. Executable specifications provide a practical representation of such expectations.

Our position is that executable specifications should serve as first-class IV\&V artifacts for QIML. We use the term to denote explicit behavioral requirements that can be assessed using suitable verification or validation mechanisms. Preconditions, postconditions, and invariants from Design by Contract (DbC) provide one practical vocabulary for expressing such requirements, but our position is not tied to a particular specification language or verifier. The central challenge is determining how to derive, represent, check, and validate these requirements in realistic QIML systems. Existing methods do not transfer directly because QIML provides limited evidence for deriving generalizable specifications, exhibits probabilistic and cross-stage behavior, spans hybrid repository-scale workflows, and often lacks strong behavioral oracles. These challenges motivate the specification-driven IV\&V perspective developed in the remainder of the paper.
\section{Executable Specifications as IV\&V Artifacts}
\label{sec:position}

IV\&V requires a behavioral reference against which an implementation can be assessed. Executable specifications can provide this reference for QIML by specifying verifiable conditions on inputs, measurement outcomes, and relationships across hybrid workflows. Quantum DbC frameworks demonstrate specifications over circuit states, measurements, and statistical post-processing~\cite{yamaguchi2023design,yamaguchi2025practical}, while deep-learning contracts extend beyond function signatures to data, model, and execution properties~\cite{shibbir2023dlcontract,dlpreconditions}. For QIML, such a scope is necessary because relevant behavior can span classical and quantum stages.

A specification states expected behavior, while verification provides evidence that it holds. Static analysis can reason about program structure~\cite{lintq}, symbolic execution can search for counterexamples~\cite{specpylot, autorespec, ayon2026autojml}, quantum-specific testing can exercise relevant behaviors~\cite{quratest}, and metamorphic testing can help when direct output oracles are unavailable~\cite{morphq}. None of these mechanisms defines intended behavior by itself; IV\&V therefore requires explicit behavioral claims assessed with evidence appropriate to each claim.

Quantum-state properties further complicate runtime checking. Measuring an intermediate state can collapse it, while the no-cloning principle prevents the retention of an arbitrary copy for independent inspection. Statistical assertions can assess expected properties over repeated measurements~\cite{huang2019statistical}. Projection-based assertions instead use projective measurements that preserve a satisfying state and can be transformed for execution on measurement-restricted quantum hardware~\cite{li2020projection}. Thus, for quantum-state properties, the checking mechanism and execution context form part of the evidence scope.
\section{Executable Specifications in QIML}
\label{sec:examples}

What should an executable specification express in a QIML application? Existing work suggests that relevant properties can occur at several points in a hybrid quantum--classical workflow. Rather than proposing a complete taxonomy, we use four representative classes to illustrate how such properties differ in scope and how they may be assessed.

\begin{table*}[t]
\centering
\small
\begin{tabular}{p{0.15\textwidth} p{0.34\textwidth} p{0.18\textwidth} p{0.21\textwidth}}

\toprule
\textbf{QIML Stage} & \textbf{Illustrative Behavioral Obligation} & \textbf{Specification Form} & \textbf{Possible Evidence} \\
\midrule

Data / parameters &
Input data and parameter dimensions are compatible with the parameterized quantum circuit &
Precondition &
Classical runtime checking on simulator or hardware \\

Circuit / state &
Circuit construction or quantum state satisfies an expected structural or state property &
Precondition / postcondition &
Simulator state inspection or hardware-feasible projection assertions \\

Measurement / post-processing &
Measurement-derived results satisfy an expected property during classical post-processing &
Postcondition &
Finite-shot execution or testing on simulator or hardware \\

Cross-stage workflow &
A condition established in one stage remains valid when consumed by a later classical or quantum stage &
Precondition / postcondition / invariant &
Integration testing or combined analysis, depending on evidence scope \\

\bottomrule
\end{tabular}
\caption{Representative executable specification obligations in a QIML workflow. Different classes of properties may require different forms of evidence for assessment.}
\label{tab:qiml_specs}
\end{table*}

\paragraph{Data and Parameter Preconditions.}
QIML components can impose relationships between classical inputs and parameterized quantum circuits. For example, quantum neural networks distinguish parameters representing input data from trainable weights, and their execution depends on inputs and weights having dimensions compatible with the corresponding circuit parameters. These conditions can be expressed as preconditions on the data and parameter configuration. Similar concerns arise in deep-learning software, where contracts may involve data and model properties that are not captured by ordinary function signatures~\cite{shibbir2023dlcontract,dlpreconditions}.

\paragraph{Circuit and State Conditions.}
Some requirements concern the construction or state of the quantum computation itself. Circuit components such as feature maps and ans\"atze must be composed consistently, and relevant properties may depend on the circuit produced through a sequence of construction operations rather than on a single function call. Quantum DbC frameworks demonstrate that assertions can be written over the input and output states of quantum circuits and checked during simulator-based testing~\cite{yamaguchi2023design,yamaguchi2025practical}. This makes circuit- and state-level properties a distinct target for executable specification.

\paragraph{Measurement and Post-Processing Conditions.}
Other properties become observable only after quantum execution. Measurement results are consumed by classical post-processing and may therefore be subject to conditions that cannot be evaluated before the circuit is run. Yamaguchi et al. explicitly support assertions over the post-processing of measurement results~\cite{yamaguchi2025practical}, while QuraTest demonstrates that quantum-specific properties such as magnitude, phase, and entanglement can guide testing~\cite{quratest}. These examples show that executable QIML specifications may need to describe properties of observed quantum behavior rather than only classical inputs and outputs.

\paragraph{Cross-Stage Conditions.} Finally, a relevant property may span more than one stage of a QIML workflow. Deep-learning contract research already shows that useful conditions can involve multiple APIs at different stages of an ML pipeline, as well as intermediate execution properties~\cite{shibbir2023dlcontract}. QIML introduces an additional boundary between classical processing and quantum computation: data and parameters configure a circuit, quantum execution produces measurements, and those results are subsequently interpreted by classical software. Learning-specific obligations can also arise during variational optimization, where finite-shot noise affects gradient estimation under both parameter-shift and finite-difference methods~\cite{mari2021estimating,piskor2022gradient}. An important question for QIML specification is, therefore, how to express and check properties whose assumptions are established in one stage but whose consequences become observable in another.
\section{Open Research Challenges}
\label{sec:challenges}

\paragraph{Limited Evidence for Generalizable Specifications.} Specification inference often depends on recurring evidence across large corpora. Nguyen et al. mined API preconditions from almost 120 million lines of Java by exploiting repeated guard conditions~\cite{nguyen2014preconditions}, while DLContract drew on documentation, prior studies, and buggy and correct deep-learning programs~\cite{shibbir2023dlcontract}. In our preliminary search for repositories using \texttt{qiskit-machine-learning} and its major submodules, after applying repository deduplication, a 10-star threshold, and exclusions for forks and archived repositories, we obtained 44 candidate repositories. These repositories contain a median of 5.1 KLOC of Python code (IQR: 1.8--11.8 KLOC). This single-framework, restricted evidence base makes recurring behavioral patterns harder to identify and raises concerns about representativeness~\cite {dabic2021sampling,borges2018githubstar}.

\paragraph{QIML Properties Span Workflow Stages.} QIML conditions may be established in one stage and observed in another. Data preparation can constrain circuit parameters, circuit construction can determine later behavior, and measurement results can affect downstream classical processing. Some properties may also require repeated observations rather than deterministic predicates. Existing quantum DbC work supports assertions over states, measurements, and post-processing~\cite{yamaguchi2023design,yamaguchi2025practical}, but representing cross-stage QIML properties remains an open problem.

\paragraph{Repository-Scale Verification Remains Difficult.} Executable specifications are useful only if they can be checked. SpecPylot uses symbolic execution for Python contracts but does not model quantum operations or quantum-state semantics~\cite{specpylot}; SpecGen integrates formal verification~\cite{specgen}, and Fun2spec validates specifications through executable assertions and tests~\cite{fun2spec}. QIML adds circuit construction, quantum operations, measurements, and classical--quantum dependencies that general-purpose symbolic tools may not model directly. Future work must determine which properties admit conventional verification and which require quantum-specific analysis, simulation, or execution.

\paragraph{A Checkable Specification May Still Be Wrong.} Executability does not guarantee that a specification captures intended behavior. Existing systems therefore use counterexamples, formal verification, tests, and manual assessment~\cite{specpylot,specgen,fun2spec}. Quantum testing similarly relies on alternative oracles when direct expected outputs are unavailable~\cite{morphq,quratest}. QIML IV\&V, therefore, needs methods that validate both implementation conformance and the credibility of the specification itself.
\section{Evidence Requirements for IV\&V}
\label{sec:evidence}

Specification-driven IV\&V should distinguish three dimensions: specification justification, implementation conformance, and evidence scope.

\paragraph{Specification Justification.}
A specification must be supported by a credible source of intent, such as an API requirement, mathematical constraint, framework semantics, or developer requirement. Executability alone is insufficient: a condition can be checkable but still encode the wrong behavior. Existing specification-generation work therefore uses verification, counterexamples, tests, or manual assessment to evaluate generated specifications~\cite{specgen,specpylot,fun2spec}.

\paragraph{Implementation Conformance.}
The next question is whether the implementation satisfies the stated property. Different properties require different evidence: static analysis can check structural properties~\cite{lintq}, symbolic execution can search for counterexamples~\cite{specpylot}, simulator-based assertions can check selected quantum-state and measurement properties~\cite{yamaguchi2025practical}, and quantum-specific or metamorphic testing can provide evidence when direct output oracles are unavailable~\cite{quratest,morphq}. A successful check supports conformance only within the scope explored.

\paragraph{Evidence Scope.}
Finally, IV\&V must state where the evidence applies. Results obtained from symbolic exploration, simulation, finite-shot execution, or a particular backend do not automatically generalize beyond those conditions. Each result should therefore record the relevant assumptions, execution context, and coverage.

Taken together, specification justification, implementation conformance, and evidence scope should remain distinct. A successful check in one dimension should not be interpreted as establishing the others.
\section{What Specification Driven\\ QIML IV\&V Requires}
\label{sec:vision}

The challenges above point to four research directions. First, QIML needs representative corpora and benchmarks for specification inference and assessment. Second, specification representations must capture cross-stage classical--quantum properties and, where needed, statistical rather than purely Boolean conditions. Third, checking mechanisms should be matched to the property being assessed, since static analysis, symbolic reasoning, simulation, testing, and execution provide different forms of evidence~\cite{lintq,morphq,quratest,yamaguchi2025practical}. Finally, QIML needs evaluation protocols that assess whether specifications are meaningful, checkable, and behaviorally discriminative. These capabilities should complement quantum benchmarking and hardware-level assessment by making behavioral claims, supporting evidence, and the scope of evidence explicit.

\subsection{Evaluation Criteria}

Future QIML specification methods should be evaluated on more than the number of conditions they generate or execute. We propose four criteria.

\paragraph{Semantic Support.}
A specification should be supported by identifiable evidence about intended behavior rather than by syntactic validity or plausibility alone. This is especially important when no reference specification is available.

\paragraph{Checkability.}
The evaluation should report which specifications can be assessed and how. Some properties may admit direct assertions or symbolic reasoning, whereas others may require simulation, repeated execution, or domain-specific analysis~\cite{specgen,specpylot,yamaguchi2025practical}.

\paragraph{Behavioral Discrimination.} A useful specification should accept behavior consistent with its intended requirement while detecting relevant violations. Evaluation should therefore consider both overly restrictive specifications and specifications too weak to expose incorrect behavior. Mutation analysis can operationalize this distinction by introducing controlled behavioral faults and measuring whether the specification detects the resulting violations. Counterexample search, testing, metamorphic relations, and domain-specific test generation provide complementary ways to assess the same distinction~\cite{specpylot,specgen,morphq,quratest}.

\paragraph{Scope and Compositionality.}
Evaluation should report where a specification applies and whether locally checked properties can support claims across multiple workflow stages. This matters because QIML behavior may depend on relationships among data preparation, circuit construction, measurement, and downstream classical processing. Prior deep-learning contract work similarly shows that relevant conditions can span multiple APIs and stages of an ML pipeline~\cite{shibbir2023dlcontract}. Studies should therefore report both the covered execution contexts and whether assumptions and guarantees can be related across component boundaries.
\section{Conclusion}
\label{conclusion}

QIML does not lack mechanisms for checking software behavior; it lacks a common layer of explicit behavioral claims around which those mechanisms can be interpreted. We argue that executable specifications can serve as first-class IV\&V artifacts. Making this practical requires credible evidence for deriving specifications, representations that capture properties spanning classical and quantum stages, checking mechanisms matched to those properties, and validation methods that distinguish implementation conformance from the correctness of the specification itself. These requirements are particularly important when available evidence is limited, relevant behavior is cross-stage or probabilistic, and direct behavioral oracles are unavailable.

The goal is not a universal verifier for QIML. Rather, specification-driven IV\&V should make each behavioral claim explicit, associate it with appropriate evidence, and state the conditions under which that evidence applies. Such an approach would allow static analysis, symbolic reasoning, simulation, testing, and execution-based evidence to contribute complementary assurance without treating any single mechanism as a complete correctness oracle. Executable specifications can therefore provide a practical foundation for making QIML behavioral claims independently assessable, comparable, and challengeable.

\bibliography{bibliography}

\end{document}